# Generalized Lorenz Systems Family

**Guanrong Chen**

Department of Electrical Engineering

City University of Hong Kong, Hong Kong SAR, China

eegchen@cityu.edu.hk

2020-6-20

**Abstract:** This article briefly introduces the generalized Lorenz systems family, which includes the classical Lorenz system and the relatively new Chen system as special cases, with infinitely many related but not topologically equivalent chaotic systems in between.

## 1. Chen System

The classical Lorenz system was coined in 1963 [1], which is described by

$$\begin{cases} \dot{x} = a(y-x), \\ \dot{y} = cx - xz - y, \\ \dot{z} = xy - bz, \end{cases} \tag{1}$$

where $a, b, c$ are real parameters. When $a = 10,\ b = \frac{8}{3},\ c = 28$, the system is chaotic, with the attractor as shown in Fig. 1 (a).

The Lorenz system has been extensively studied in, for example, chaos theory, bifurcation analysis, as well as chaos control and synchronization [2-4]. To this end, one might wonder if the Lorenz system is just a stand-alone lucky discovery, or if there are other closely-related chaotic systems of the same kind? This short note summarizes some recent findings and research progress on this concerned topic.

First, recall a couple of historical instances. From the so-called “anti-control” approach, known also as chaotification [5,6], Chen found a new system [7], referred to as the Chen system lately [8,9], described by

$$\begin{cases} \dot{x} = a(y - x), \\ \dot{y} = (c - a)x - xz + cy, \\ \dot{z} = xy - bz, \end{cases} \tag{2}$$

where $a, b, c$ are real parameters. When $a = 35,\ b = 3,\ c = 28$, the system is chaotic, with the attractor as shown in Fig. 1 (b).

The anti-control approach here is adding to the right-hand side of the second equation of the Lorenz system, which governs the most vigorous dynamics, the following linear controller:

$$u = \alpha x + \beta y + \gamma z\,,$$

where $\alpha, \beta, \gamma$ are real constants to be determined. Then, under the Shilnikov condition [10], it determines the coefficients $\alpha,\ \beta,\ \gamma$ such that the resultant controlled system is chaotic. It turned out that one simple choice is $\alpha = -a,\ \beta = c + 1,\ \gamma = 0$, yielding the new system (2).

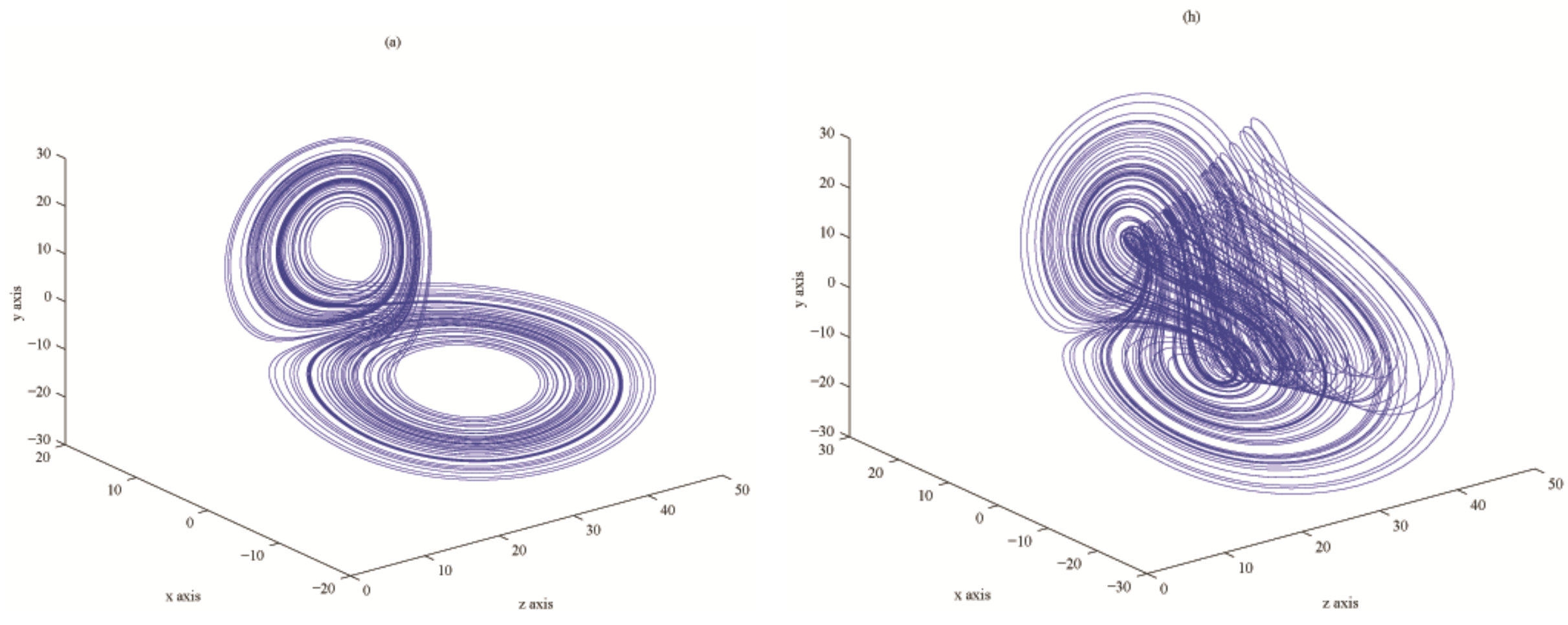

**Fig. 1** Phase portraits of the two systems: (a) Lorenz attractor; (b) Chen attractor [4]

The Chen system (2) appears to be quite similar to the Lorenz system (1), as it is supposed to be since system (2) comes from system (1). Thus, due to their similar structures, they must be closely related and should also have similar properties.

As such, does the Chen system have anything new, namely is it worth further investigating?

To answer this question, it needs to relate and also distinguish these two systems from a dynamics point of view.

Naturally, the first concern is: are these two systems actually equivalent?

In dynamical systems theory, two systems are smoothly state equivalent if and only if there exists a diffeomorphism $f(x, y, z) \to (u, v, w)$ transforming the states of one system to that of the other. If so, then except for the appearance of their formulations the two systems have no essential differences regarding their dynamical properties, therefore can be considered to be the same system. Yet, the answer to this concerned question has been confirmed: it is proved in [11] that the Chen system (2) and the Lorenz system (1) are not smoothly state equivalent.

Here, it should be noted that in performing any state transformation on an autonomous system, the time variable $t \in [0, \infty)$ will not be transformed, especially will not be transformed reversely: $t \to \tau = -t$. Otherwise, many theoretical and technical problems will become meaningless. For example, a stable system with a solution converging to zero as $t \to +\infty$ will become unstable with a solution diverging to infinity as $\tau \to -\infty$, therefore cannot be analyzed (for instance, the former has a bounded invariant $\omega$-limit set but the latter does not, which cannot even be well defined or is trivial).

Of course, one may bypass the autonomous systems theory to perform a state transformation with time reversal $t \to \tau = -t$, to see if there is anything new? To proceed, consider a sample case [12]. In the Lorenz system (1), let $a = 0.4,\ b = -1.4,\ c = -0.4$. Then, one has a dimensionless system

$$\begin{cases} \dot{x} = 0.4(y - x), \\ \dot{y} = -1.4x - xz - y, \\ \dot{z} = xy + 0.4z. \end{cases} \tag{3}$$

Similarly, in the Chen system (2), letting $a = -0.4,\ b = 0.4,\ c = 1.0$ gives

$$\begin{cases} \dot{x} = -0.4(y - x), \\ \dot{y} = 1.4x - xz + y, \\ \dot{z} = xy - 0.4z. \end{cases} \tag{4}$$

Now, in system (3) let $t \to +\infty$, meanwhile in system (4) let $t \to -\infty$. One will find that the orbits of the two systems converge to the same bounded limit set, which is a chaotic attractor, as shown in Fig. 2.

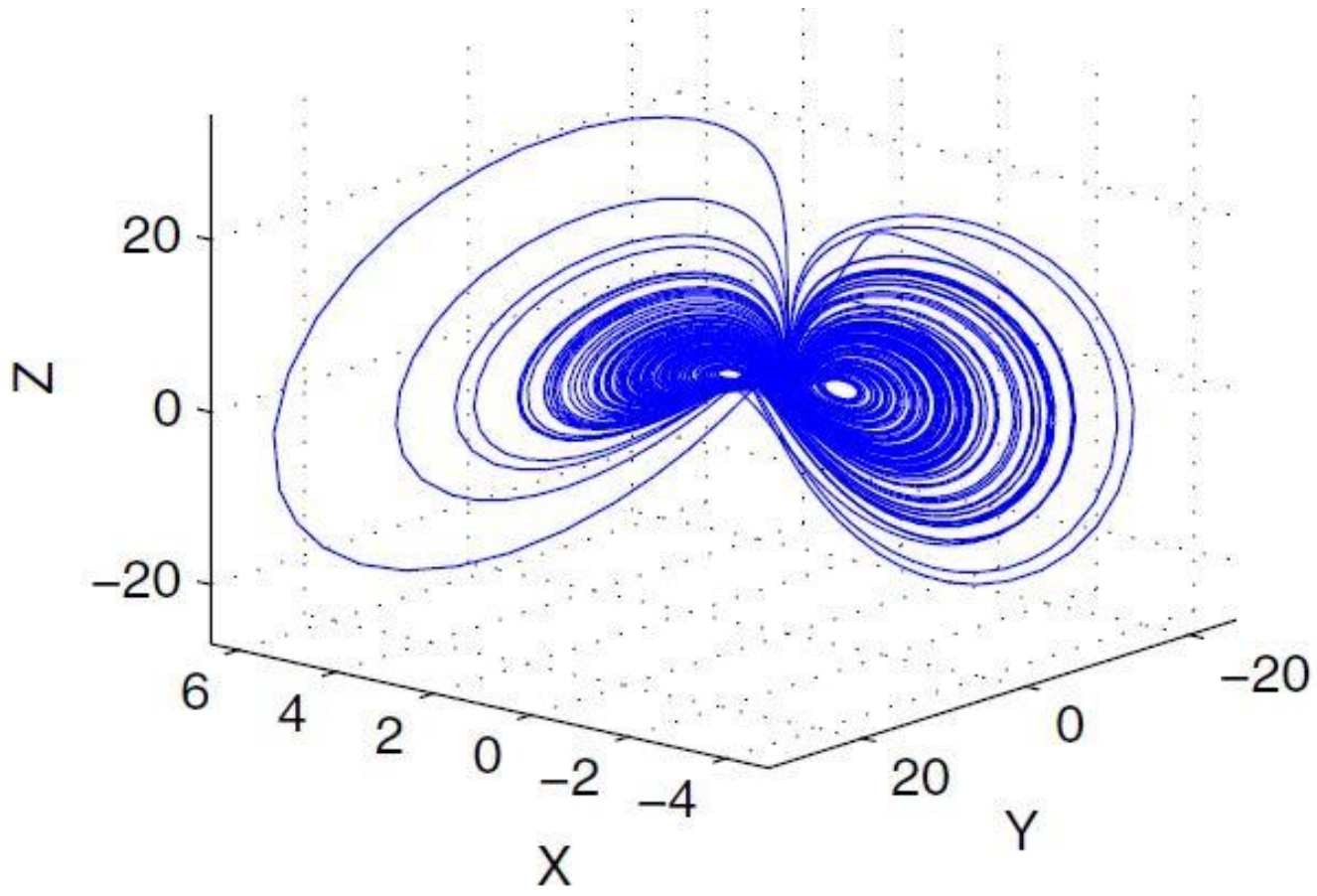

**Fig. 2** Common limit set of system (3) as $t \to +\infty$ and system (4) as $t \to -\infty$, which is a chaotic attractor [12]

This example shows that the two systems "shake hands" at opposite infinity time limits. It implies that, on the one hand, the two systems are closely related, while on the other hand, they are very different: if one lets $t \to +\infty$ in the Lorenz system (1) and let $t \to -\infty$ in the same system or any smoothly state equivalent system, the results will be totally different (one converges and the other divergens) and they never "shake hands" anywhere.

Two non-equivalent systems naturally have different dynamical behaviors, and yet the Chen system (2) and the Lorenz system (1) have the aforementioned relationships, so the next question is what kinds of different dynamical behaviors they may have, or whether there is a need to study the Chen system given the existing knowledge of the Lorenz system.

At first glance, it can be clearly seen from Figs. 1 (a)-(b) that the Chen attractor is more complicated than the Lorenz attractor, at least visually. In fact, the former is more complex than the latter in terms of chaotic and bifurcative dynamics, as well as multistability, as shown by Fig. 3 in comparing their stable manifolds [9], by Fig. 4 in comparing their forward and backward invariant limit sets [13], and by Fig. 5 in comparing their topological foliations [14].

It is found [15] that, compared to the Lorenz system (1), the first return map of the Chen system (2) exhibits a more complex structure of continuous regions and there are six crossing blocks with respect to the second return map.

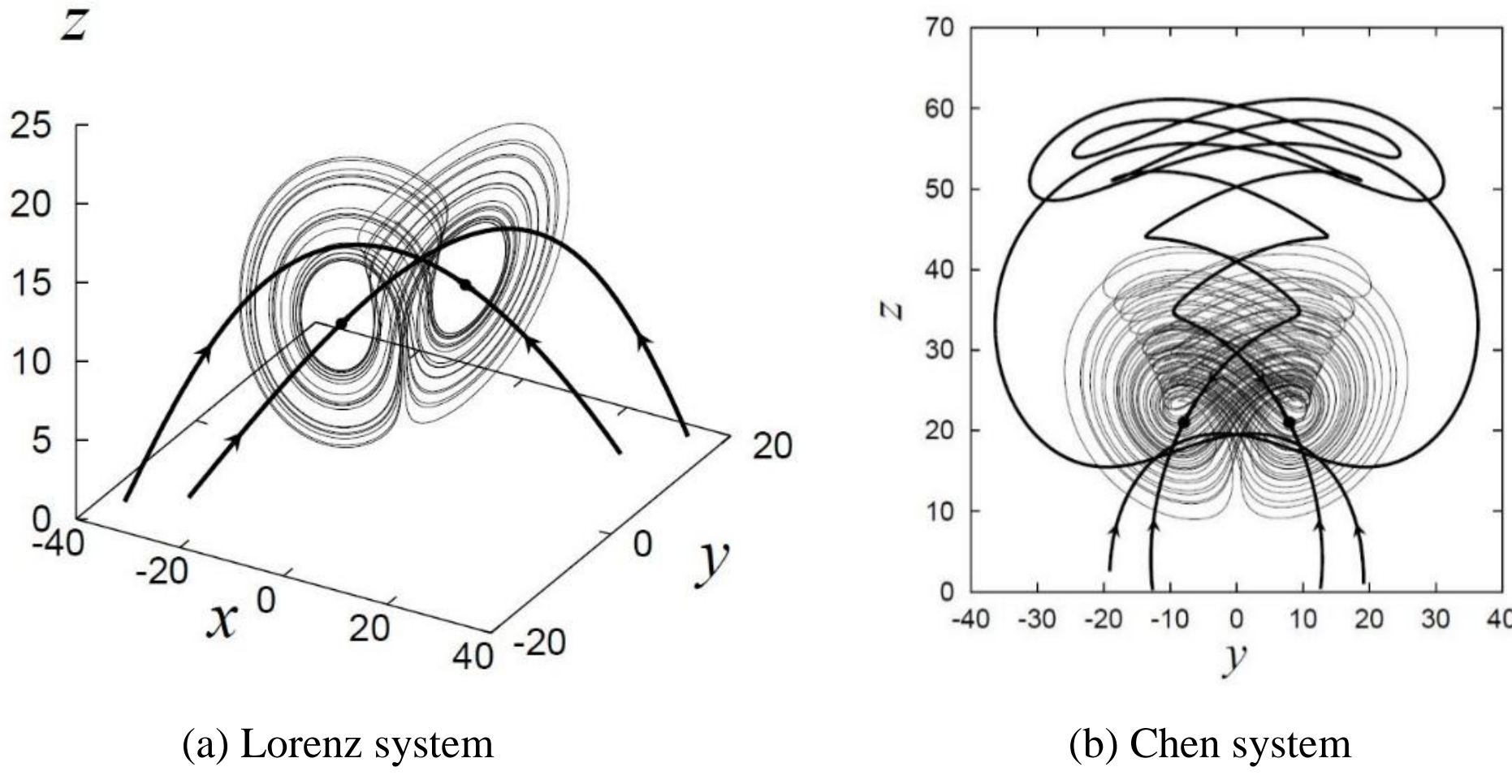

(a) Lorenz system (b) Chen system

**Fig. 3** Stable manifolds of the two systems [9]

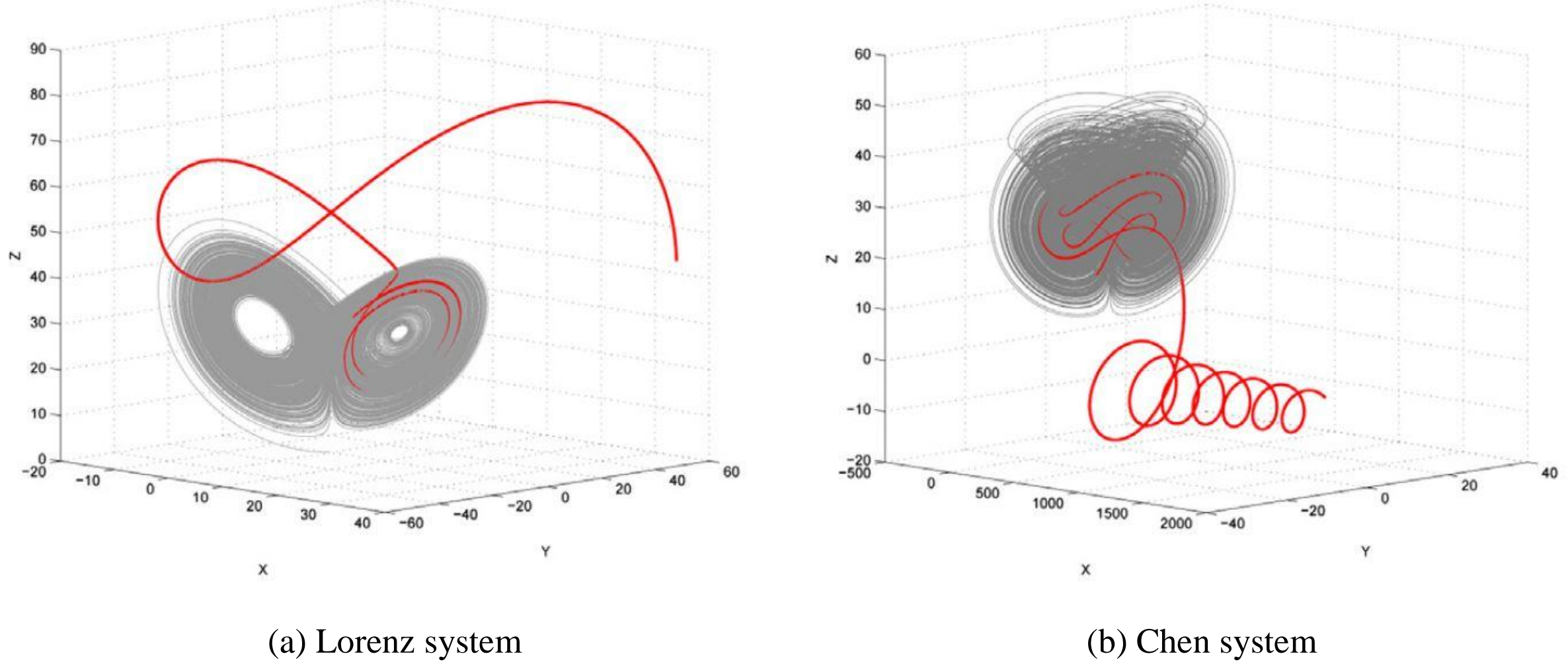

(a) Lorenz system (b) Chen system

**Fig. 4** Forward and backward invariant sets of the two systems [13]

(gray is forward invariant set, as $t \to +\infty$; red is backward invariant set, as $t \to -\infty$)

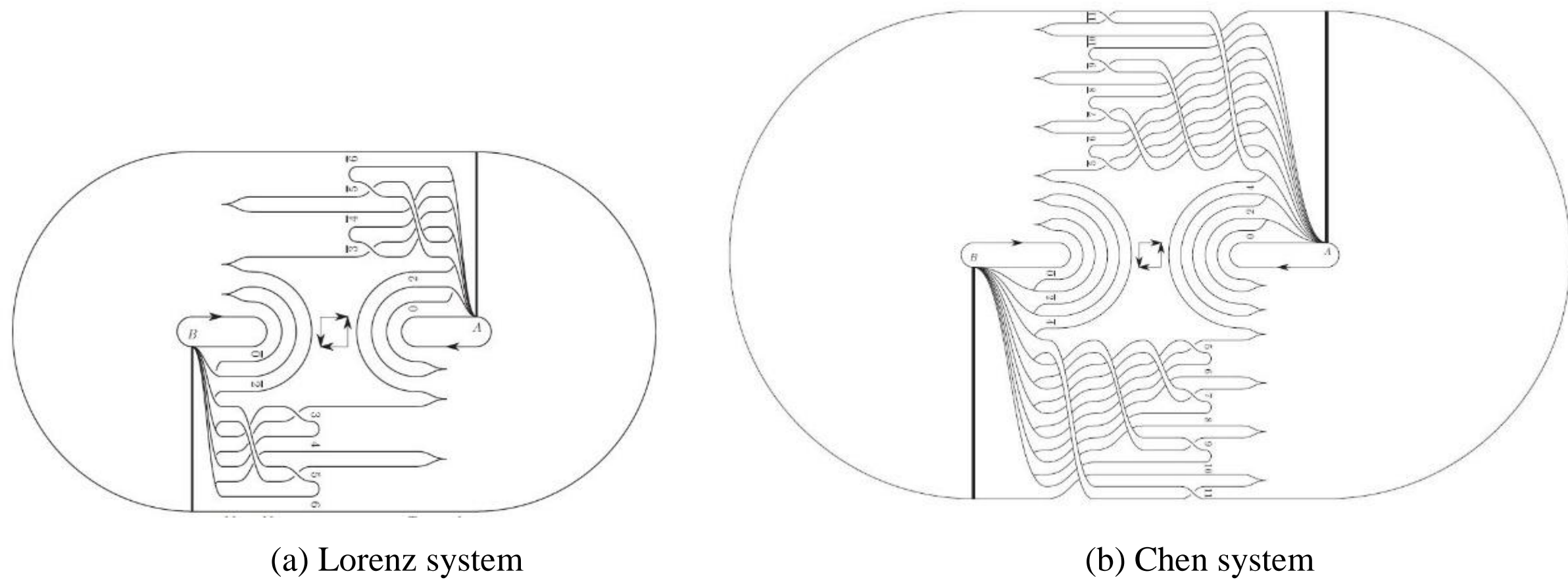

(a) Lorenz system (b) Chen system

**Fig. 5** Topological foliations of the two systems [14]

Second, the Lorenz system is globally ultimately bounded in the sense that, starting from anywhere in the phase space, the system orbit will eventually fall into the following spherical surface and be trapped inside thereafter [16]:

$$x^2 + y^2 + (z - a - c)^2 = \frac{b^2(a+c)^2}{4(b-1)},$$

where $a > 0,\ b > 1,\ c > 0$. However, for the Chen system, no such closed-form surface expression has been found so far, causing a lot of complication and difficulties in theoretical analysis of the Chen system. The problem comes from the two sign-opposite terms in the second equations of the two systems, $-y$ and $+cy$ with $c > 0$, where the latter leads to some kind of divergence causing many technical troubles for analysis. Nevertheless, the global ultimate boundedness of the Chen system can still be proved, but using different techniques based on Lyapunov functions and under different conditions such as $a \geq b + c,\ b > 0,\ c > 0$ in [17], or in some other regions as reported elsewhere.

Furthermore, regarding the dissipativeness it is easy to verify that the divergence of the Lorenz system is div(L) $= -(a + b + 1)$, which is alway negative for $a + b > -1$ and this condition is independent of the parameter $c$. Since in the chaotic regime of the Lorenz system, $a, b > 0$, the above condition is always satisfied. So, the chaotic Lorenz system is always dissipative. However, the divergence of the Chen system (2) is div(C) $= -(a + b - c)$, which is negative only on one side of the plane $a + b = c$, thus the dissipativeness of the Chen system is determined by the parameter $c$.

To proceed looking at bifurcations, the Lorenz system (1) has a key bifurcation parameter [2]

$$c_0 = (a + b + 3)/(a - b - 1),$$

while the Chen system (2) has the corresponding parameter [18]

$$\mathrm{c}_0 = \frac{1}{2}\left(\sqrt{17a^2 - 6ab + b^2} - 3a + b\right).$$

It is shown in [19] that the Hopf bifurcation surfaces of the two systems fall respectively on the oppose sides of the parameter region:

Lorenz system: $c_0 > a + b$,

Chen system: $c_0 < a + b$.

To move further on, consider the Lyapunov dimensions of the two systems. Lorenz system (1) has the Lyapunov dimension [20]

$$\dim_L K = 3 - \frac{2(a+b+1)}{a+1+\sqrt{(a-1)^2+4ac}}.$$

However, the Lyapunov dimension formula for the Chen system (2) still remains open today.

Next, it is important to compare the independent parameters of the two systems. For the Chen system (2), the following state and time transformations [20]

$$u = \frac{x}{b}, \quad v = \frac{y}{b}, \quad w = \frac{z}{b}, \quad \tau = bt \,, \tag{5}$$

change the Chen system (2) to

$$\begin{cases} \dot{u} = \alpha(v - u), \\ \dot{v} = (\gamma - \alpha)u - uw + \gamma v, \\ \dot{w} = uv - w, \end{cases} \tag{6}$$

where $\alpha = a/b$, $\gamma = \frac{c}{b}$, thus the parameter $b$ disappeared. This means that, by nature, the Chen system has only 2 independent parameters. With one less parameter but having more complicated behaviors than the Lorenz system, this shows a certain advantage of the Chen system in the study of complex system dynamics. Moreover, this demonstrates some essential differences betwenn the two systems in terms of both topological and dynamical characteristics.

Forthermore, to reveal the structural properties of all Lorenz types of systems, a universal normal form of the generalized Lorenz system is proposed in [21]:

$$\begin{cases} \dot{x} = y, \\ \dot{y} = -(x^2 + z - 1)x - \lambda y, \\ \dot{z} = -\alpha z + \beta x^2, \end{cases}$$

with positive parameters $\alpha, \beta, \gamma$. The paper demonstrates the forming of many complex orientable and non-orientable attractors, including the Lorenz and Chen attractors and their like attractors. A key role in such a general formation is played by the homoclinic orbits of the saddle, twisted around the axis of symmetry, which were referred to as the twisted homoclinic orbits. The paper shows that decreasing the parameter value of λ causes an intricacy of the universal normal form bifurcation set, whereas the dynamics in the parameter zone of the Chen system are more complex than that in the parameter zone of the Lorenz system [21].

Finally, it is intersting to consider the physics behind the two systems. Lorenz system (1) has a familiar physical background, which is a simplified weather dynamics model, where $x$ is a variable of the spatial average of hydrodynamic velocity, $y$ is a variable of temperature difference between the ascending and descending currents, $z$ is a variable of the temperature gradient proportional to the distortion of the vertical temperature profile from linearity, $a$ is the Prandtl constant, $b$ is a proportional constant, and $c$ is the Rayleigh number [1,2]. The anti-controller used to generate the Chen system is given, as shown above, by

$$u = \alpha x + \beta y + \gamma z = -ax + (c+1)y + 0z,$$

where the first term adds some negative velocity while the second term adds some positive heat flux to the original Lorenz system. It can be seen that, in this anti-controller, the first term regulates the flow velocity while the second term changes the negative heat flux $-y$ of the Lorenz system to be a positive heat flux $+cy$. This creates a shear force changing the free flow to be a forced one, which eventually yields new turbulence into the Lorenz system. The joint effect of these two additional terms leads the flow to move rapidly along the vertical direction, as can be clearly seen from athe comparison of the Lorenz attractor shown in Fig. 1(a) and the Chen attractor shown in Fig. 1(b). Thus, the Chen system can be interpreted physically as a temperature-controlled Lorenz weather system [22].

The first physical circuit of the Chen system was constructed in [23] and a fully integrated circuit of the Chen system was designed and fablicated in [24].

It should be pointed out that the significance of the Chen system (2) has gone much beyond the discovery of this system itself. Indeed, it has induced a new generalized Lorenz systems family and many Lorenz-like systems [25] such as the T-system [26], and led to some new research problems and methodologies [13].

In the following, the generalized Lorenz systems family is briefly introduced along with its canonical form.

## 2. Generalized Lorenz systems family and its canonical form

First, recall a Lorenz system canonical form introduced by Celikovsky and Vanecek [27], which separates the linear part and nonlinear part of the Lorenz system:

$$\begin{bmatrix} \dot{x} \\ \dot{y} \\ \dot{z} \end{bmatrix} = \begin{bmatrix} a_{11} & a_{12} & 0 \\ a_{21} & a_{22} & 0 \\ 0 & 0 & a_{33} \end{bmatrix} \begin{bmatrix} x \\ y \\ z \end{bmatrix} + x \begin{bmatrix} 0 & 0 & 0 \\ 0 & 0 & -1 \\ 0 & 1 & 0 \end{bmatrix} \begin{bmatrix} x \\ y \\ z \end{bmatrix}. \qquad (7)$$

It was noted that the Lorenz system (1) satisfies $a_{12}a_{21} > 0$, and it is easy to see that the Chen system (2) satisfies $a_{12}a_{21} < 0$. In this sense, the two systems are dual to each other.

In 2002, Lü and Chen [28] discovered a chaotic system, called the Lü system lately, which satisfies $a_{12}a_{21} = 0$, representing a transition between the Lorenz system and the Chen system:

$$\begin{cases} \dot{x} = a(y - x), \\ \dot{y} = -xz + cy, \\ \dot{z} = xy - bz. \end{cases} \tag{8}$$

When $a = 36,\ b = 3,\ c = 20$ , this system is chaotic with the attractor as shown in Fig. 6.

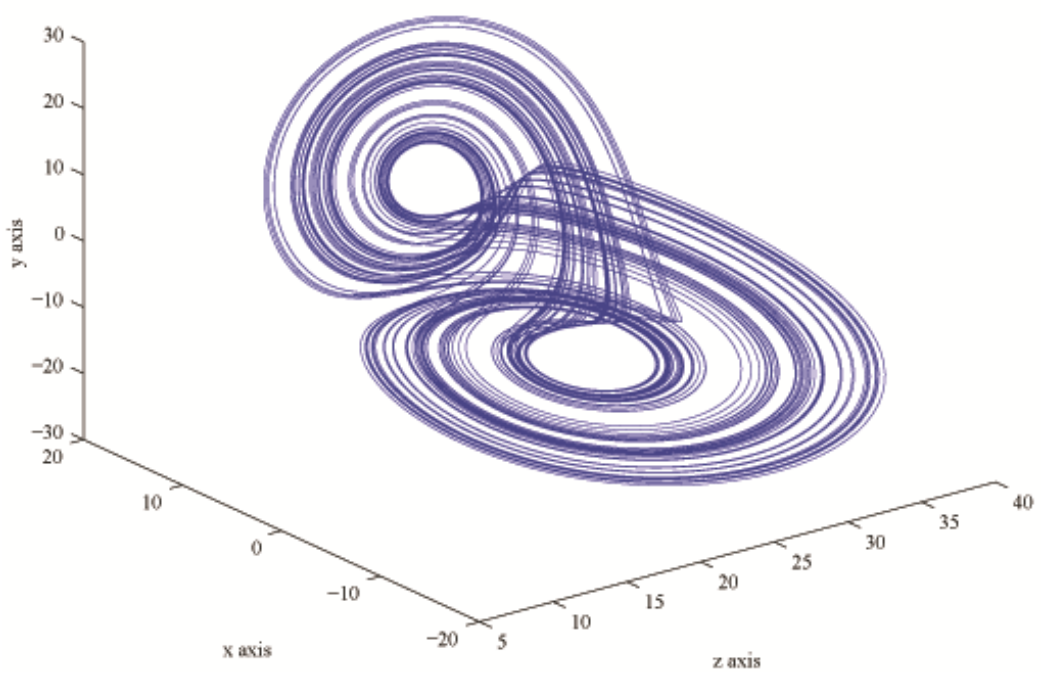

**Fig. 6** Lü attractor [25]

Consequently, Lü et al. [29] constructed a unified system, which connects the Lorenz system, Lü system and Chen system together:

$$\begin{cases} \dot{x} = (25\alpha + 10)(y - x), \\ \dot{y} = (28 - 35\alpha)x - xz + (29\alpha - 1)y, \\ \dot{z} = xy - \frac{1}{3}(\alpha + 8)z, \end{cases} \tag{9}$$

where $\alpha$ is a real parameter. It can be numerically verified that, for any $\alpha \in [0, 1]$, system (9) is chaotic.

This unified system by nature is a convex combination of the Lorenz system and the Chen system, but it represents infinitely many chaotic systems in between, as a family of chaotic systems with the same structure. The Lorenz system and the Chen system are two extremes: when $\alpha = 0$, it is the Lorenz system; when $\alpha = 1$, it is the Chen system; when $\alpha$ is varied from 0 to 1, all systems are chaotic as can be verified numerically.

To this end, it can be realized that the Lorenz system is not a stand-alone instance, but is associated with many closely-related chaotic systems. As one more example, lately Yang and Chen [30] constructed the following system, referred to as the Yang system [31]:

$$\begin{cases} \dot{x} = a(y - x), \\ \dot{y} = cx - xz, \\ \dot{z} = xy - bz. \end{cases} \tag{10}$$

When $a = 10,\ b = \frac{8}{3},\ c = 16$, this system is chaotic with the attractor as shown in Fig. 7.

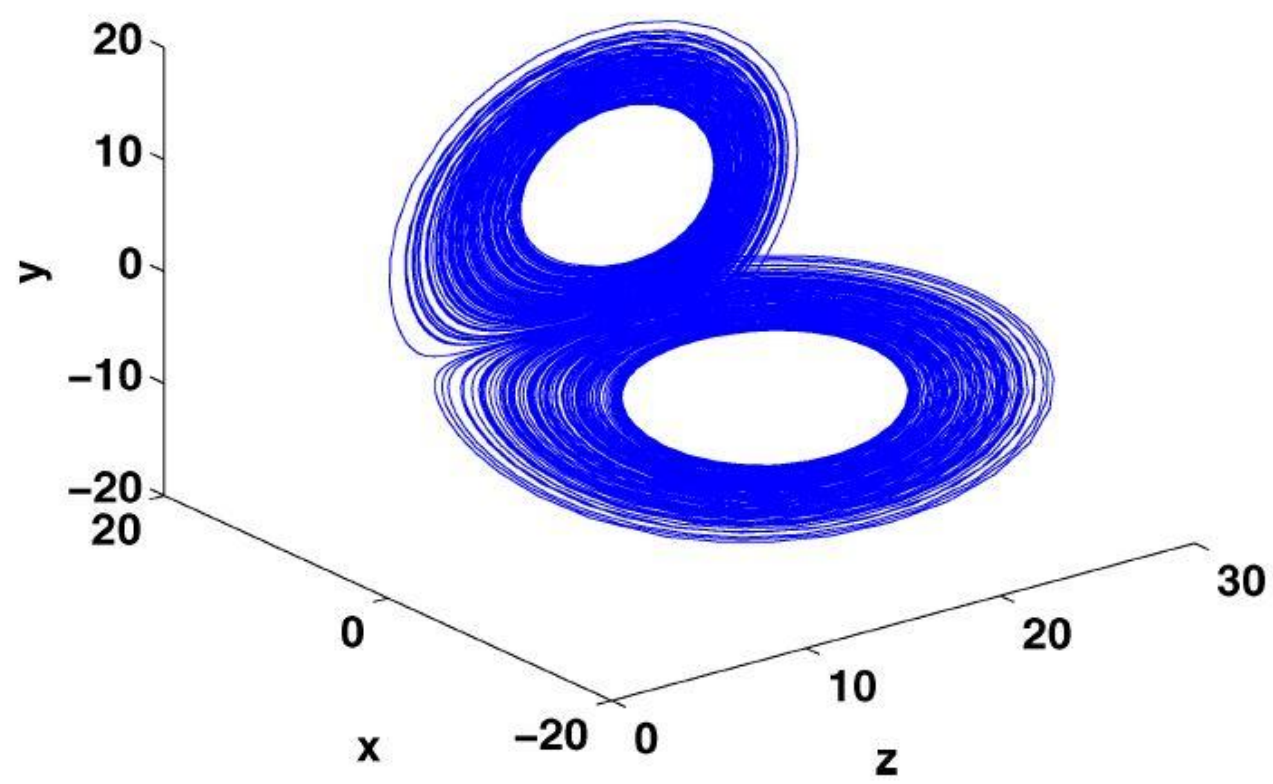

**Fig. 7** Yang attractor [27]

In system (7), by looking at $a_{11}a_{22}$ rather than $a_{12}a_{21}$, the Lorenz system (1) satisfies $a_{11}a_{22} > 0$, the Chen system (2) satisfies $a_{11}a_{22} < 0$, while the Yang system (10) satisfies $a_{11}a_{22} = 0$. So, in this sense, the Chen system and the Lorenz system are also dual to each other.

Along this line of thinking, Celikovsky and Chen [32] formulated the following generalized Lorenz systems (GLS) family:

$$\dot{x} = \begin{bmatrix} a_{11} & a_{12} & 0 \\ a_{21} & a_{22} & 0 \\ 0 & 0 & \lambda_3 \end{bmatrix} x + x_1 \begin{bmatrix} 0 & 0 & 0 \\ 0 & 0 & -1 \\ 0 & 1 & 0 \end{bmatrix} x, \tag{11}$$

where $x = [x_1,\ x_2,\ x_3]^T$ and the first matrix has real eigenvalues satisfying the Shilnikov condition: $-\lambda_2 > \lambda_1 > -\lambda_3 > 0$. It is easy to verify that this GLS contains the Lorenz, Chen, Lü and Yang systems as special cases. The GLS (11) is said to be non-trivial if it has at least one solution that does not tend to 0, or infinity, or a limit cycle.

Further, they extended the GLS family to a one-parameter generalized Lorenz canonical form (GLCF) [33]:

$$\dot{z} = \begin{bmatrix} \lambda_1 & 0 & 0 \\ 0 & \lambda_2 & 0 \\ 0 & 0 & \lambda_3 \end{bmatrix} z + (\mu z) \begin{bmatrix} 0 & 0 & -1 \\ 0 & 0 & -1 \\ 1 & \tau & 0 \end{bmatrix} z, \tag{12}$$

where $z = [z_1,\ z_2,\ z_3]^T$, $\mu = [1,\ -1,\ 0]$, $-\lambda_2 > \lambda_1 > -\lambda_3 > 0$, with parameter $\tau \in R$.

Meanwhile, they proved [33] that there exists a nonsingular one-parameter linear state transformation $z = T_\tau x$ converting a non-trivial GLS (11) to GLCF (12) with $\tau > -1$.

Then, they considered the case of $\tau \le -1$, for which the GLCF (12) is actually not well defined.

First, for $\tau < -1$, they introduced the following hyperbolic generalized Lorenz system (HGLS) [34]:

$$\dot{x} = \begin{bmatrix} a_{11} & a_{12} & 0 \\ a_{21} & a_{22} & 0 \\ 0 & 0 & \lambda_3 \end{bmatrix} x + x_1 \begin{bmatrix} 0 & 0 & 0 \\ 0 & 0 & 1 \\ 0 & 1 & 0 \end{bmatrix} x, \tag{13}$$

where $x = [x_1,\ x_2,\ x_3]^T$, $-\lambda_2 > \lambda_1 > -\lambda_3 > 0$. HGLS (13) is said to be non-trivial if it has at least one solution that does not tend to 0, or infinity, or a limit cycle.

Clearly, HGLS (13) differs from GLS (11) only in the sign of a "1" inside the second matrix. However, this difference is very significant, so that they cannot be put together into one canonical form, where the second matrix of (11) has eigenvalues $\{0, \pm j\}$ while that of (13) has $\{0, \pm 1\}$; in this sense, system (13) is named as a hyperbolic ssytem.

Now, GLS (11) and HGLS (13) can be combined together, with $\tau \neq -1$, as follows:

$$\dot{x} = \begin{bmatrix} a_{11} & a_{12} & 0 \\ a_{21} & a_{22} & 0 \\ 0 & 0 & \lambda_3 \end{bmatrix} x + x_1 \begin{bmatrix} 0 & 0 & 0 \\ 0 & 0 & -\mathrm{sgn}(\tau+1) \\ 0 & 1 & 0 \end{bmatrix} x, \tag{14}$$

where $x = [x_1,\ x_2,\ x_3]^T$, $-\lambda_2 > \lambda_1 > -\lambda_3 > 0$.

To this end, they showed that for the case of $\tau = -1$, the GLCF (12) can also be well defined but in a different form [33], leading to an equivalent Shimidzu-Morioka system [35]:

$$\begin{cases} \dfrac{dx}{d\theta} = y, \\ \dfrac{dy}{d\theta} = x(1-z) - \alpha y, \\ \dfrac{dz}{d\theta} = -\beta z + x^2, \end{cases}$$

where $\theta = t\sqrt{-\lambda_1\lambda_2}$, $\alpha = -(\lambda_1 + \lambda_2)/\sqrt{-\lambda_1\lambda_2}$, $\beta = \lambda_3/\sqrt{-\lambda_1\lambda_2}$.

Note that the GLCF (12) is defined according to the algebraic structures of the systems, which includes all the systems discussed above as special cases. It has only one parameter satisfying $-\infty < \tau < +\infty$, where the Lorenz system satisfies $0 < \tau < +\infty$, the Lü system satisfies $\tau = 0$, the Chen system satisfies $-1 < \tau < 0$, the Shimidzu-Morioka system satisfies $\tau = -1$, and the hyperbolic system satisfies $-\infty < \tau < -1$. This also indicates, from yet another point of view, that the Lorenz system, Lü system and Chen system are not smoothly state equivalent to each other.

Finally, it should be mentioned that, under the condition of $-\lambda_2 > \lambda_1 > -\lambda_3 > 0$, it was proved in [36] that the GLCF is chaotic in the sense of Shilnikov.

The GLCF (12) can be used to classify a large number of chaotic systems that are closely related to the classical Lorenz system, as summarized in Table I. Detailed dynamical analysis of the GLCF was first given in [4], and a more precise verification with specific state transformations are provided in [37].

**Table I:** GLCF-equvalent systems and their classification

| GLCF | Equivalent systems and representative examples |
|---|---|
| $\tau \in (-\infty, -1)$ | Hyperbolic generalized Lorenz system HGLS |
| $\tau = -1$ | Shimidzu-Morioka system |
| $\tau \in (-1, 0)$ | Generalized Lorenz system GLS, with $a_{12}a_{21} < 0$; Chen system |
| $\tau = 0$ | Generalized Lorenz system GLS with $a_{12}a_{21} = 0$; Lü system |
| $\tau \in (0, \infty)$ | Generalized Lorenz system GLS with $a_{12}a_{21} > 0$; Lorenz system |

Along the same approach, Yang et al. [38] showed that there exists a nonsingular one-parameter linear state transformation $z = T_\tau x$ converting both (12) and (13) together to the following complete Lorenz canonical form (CLCF):

$$\dot{z} = \begin{bmatrix} \lambda_1 & 0 & 0 \\ 0 & \lambda_2 & 0 \\ 0 & 0 & \lambda_3 \end{bmatrix} z + \mu\, z \begin{bmatrix} 0 & 0 & \mathrm{sgn}(\tau) \\ 0 & 0 & \mathrm{sgn}(\tau) \\ -a_{12}^2\Delta - \xi & \xi & 0 \end{bmatrix} \qquad (15)$$

where $z = [z_1,\ z_2,\ z_3]^T$, $\mu = [1,\ -1,\ 0]$, $\xi = a_{12}^2\sqrt{\Delta}\,(a_{22} - \lambda_1)$, $\Delta = [\mathrm{tr(A)}]^2 - 4\det(\mathrm{A})$, $A = \begin{bmatrix} a_{11} & a_{12} \\ a_{21} & a_{22} \end{bmatrix}$. This system is actually characterized by two parameters, $\tau$ and $\xi$, which also contains the Lorenz, Chen, and Lü systems as special cases including their so-called conjugate counterparts. The CLCF (15) is said to be non-trivial if it has at least one solution that does not tend to 0, or infinity, or a limit cycle.

Similarly, CLCF (15) can be used to classify a large number of chaotic systems that are closely related to the classical Lorenz system, as summarized in Table II [38], where the case of $\tau = 0$ is trivial therefore not listed.

**Table II:** CLCF-equivalent systems and their classification

| $\tau$ | $\xi$ | Equivalent systems | Representative examples |
|---|---|---|---|
| $\tau < 0$ | $\xi \in (-\infty, -a_{12}^2\Delta)$ | GLS | |
| | $\xi = -a_{12}^2\Delta$ | | |
| | $\xi \in (-a_{12}^2\Delta, 0)$ | | Lorenz system |
| | $\xi = 0$ | | Lü system |
| | $\xi \in (0, +\infty)$ | | Chen system |
| $\tau > 0$ | $\xi \in (-\infty, -a_{12}^2\Delta)$ | CLCF | Conjugate Chen system |
| | $\xi = -a_{12}^2\Delta$ | | Conjugate Lü system |
| | $\xi \in (-a_{12}^2\Delta, 0)$ | | Conjugate Lorenz system |
| | $\xi = 0$ | | |
| | $\xi \in (0, +\infty)$ | | |